\documentclass[12pt]{article}
\usepackage{amsmath}
\usepackage{graphicx}
\usepackage{natbib}
\usepackage{url} 
\usepackage{algorithm}
\usepackage{algpseudocode}
\usepackage{multirow}

\newcommand{\blind}{0}

\begin{document}

\def\spacingset#1{\renewcommand{\baselinestretch}%
{#1}\small\normalsize} \spacingset{1}

\newcommand{\bbeta}{{\mbox{\boldmath$\beta$}}}
\newcommand{\bEta}{{\mbox{\boldmath$\eta$}}}
\newcommand{\bdelta}{{\mbox{\boldmath$\delta$}}}
\newcommand{\bgamma}{{\mbox{\boldmath$\gamma$}}}
\newcommand{\bmu}{{\mbox{\boldmath$\mu$}}}
\newcommand{\balpha}{{\mbox{\boldmath$\alpha$}}}
\newcommand{\btheta}{{\mbox{\boldmath$\theta$}}}
\newcommand{\bTheta}{{\mbox{\boldmath$\Theta$}}}
\newcommand{\bphi}{{\mbox{\boldmath$\phi$}}}
\newcommand{\bSigma}{{\mbox{\boldmath$\Sigma$}}}
\newcommand{\bLambda}{{\mbox{\boldmath$\Lambda$}}}
\newcommand{\bXi}{{\mbox{\boldmath$\Xi$}}}
\newcommand{\bpi}{{\mbox{\boldmath$\pi$}}}
\newcommand{\bveps}{{\mbox{\boldmath$\varepsilon$}}}
\newcommand{\R}{\texttt{R}}
\newcommand{\Lik}{\mathcal{L}}
\newcommand{\bb}{\textbf{b}}
\newcommand{\bx}{\textbf{x}}
\newcommand{\bl}{\textbf{l}}
\newcommand{\by}{\textbf{y}}
\newcommand{\bz}{\textbf{z}}
\newcommand{\bu}{\textbf{u}}
\newcommand{\bg}{\textbf{g}}
\newcommand{\bw}{\textbf{w}}
\newcommand{\bX}{\textbf{X}}
\newcommand{\bZ}{\textbf{Z}}
\newcommand{\bS}{\textbf{S}}
\newcommand{\bc}{\textbf{c}}
\newcommand{\bC}{\textbf{C}}
\newcommand{\bB}{\textbf{B}}
\newcommand{\bY}{\textbf{Y}}
\newcommand{\bM}{\textbf{M}}
\newcommand{\bE}{\textbf{E}}
\newcommand{\bV}{\textbf{V}}
\newcommand{\bU}{\textbf{U}}
\newcommand{\bW}{\textbf{W}}
\newcommand{\bL}{\textbf{L}}
\newcommand{\bI}{\textbf{I}}
\newcommand{\rest}{\text{rest}}
\newcommand{\one}{\textbf{1}}
\newcommand{\zero}{\textbf{0}}
\newcommand{\up}{\underline{p}}


\if0\blind
{
  \title{\bf Model Selection in Variational Mixed Effects Models}
  \author{Mark J. Meyer\\
    Department of Mathematics and Statistics, Georgetown University\\
    and \\
    Selina Carter \\
    Department of Statistics and Data Science, Carnegie Mellon University\\
    and \\
    Elizabeth J. Malloy \\
    Department of Mathematics and Statistics, American University}
  \maketitle
} \fi

\if1\blind
{
  \bigskip
  \bigskip
  \bigskip
  \begin{center}
    {\LARGE\bf Model Selection in Variational Mixed Effects Models}
\end{center}
  \medskip
} \fi

\bigskip
\begin{abstract}
Variational inference is an alternative estimation technique for Bayesian models. Recent work shows that variational methods provide consistent estimation via efficient, deterministic algorithms. Other tools, such as model selection using variational AICs (VAIC) have been developed and studied for the linear regression case. While mixed effects models have enjoyed some study in the variational context, tools for model selection are lacking. One important feature of model selection in mixed effects models, particularly longitudinal models, is the selection of the random effects which in turn determine the covariance structure for the repeatedly sampled outcome. To address this, we derive a VAIC specifically for variational mixed effects (VME) models. We also implement a parameter-efficient VME as part of our study which reduces any general random effects structure down to a single subject-specific score. This model accommodates a wide range of random effect structures including random intercept and slope models as well as random functional effects. Our VAIC can model and perform selection on a variety of VME models including more classic longitudinal models as well as longitudinal scalar-on-function regression.  As we demonstrate empirically, our VAIC performs well in discriminating between correctly and incorrectly specified random effects structures. Finally, we illustrate the use of VAICs for VMEs on two datasets: a study of lead levels in children and a study of diffusion tensor imaging.
\end{abstract}

\noindent%
\vfill

\newpage
\spacingset{1.5} 

\section{Introduction}
\label{s:intro}

	Variational inference is an increasingly popular tool for estimating Bayesian models that relies on an approximation of the joint posterior distribution to obtain tractable solutions for the marginal posteriors of model parameters \citep{OrmerodWand2010,Blei2017}. The resulting algorithms are relatively efficient for estimating parameters. As an illustration, consider a vector of parameters, $\btheta$, and a vector of observed data, $\bY$. The posterior distribution for $\btheta$ given $\bY$ is $p(\btheta | \bY) = p(\bY, \btheta)\big/p(\bY)$. Typically, $p(\bY)$ is intractable and posterior estimates need to be obtained algorithmically. For the arbitrary density $q$, $p(\bY)$ is bounded below by $\up(\bY; q)$ where 
		$\up(\bY; q) = \exp\left[ \int q(\btheta) \log\left\{ \frac{p(\bY, \btheta)}{q(\btheta)} \right\} d\btheta \right].$ Variational algorithms work by maximizing $\up(\bY; q)$ over a class of densities, $q$, that are tractable. This in turn minimizes the Kullback-Leibler (KL) Divergence between the approximation, $q(\btheta)$, and the posterior, $p(\btheta | \bY)$.
	
	Given a partition of $\btheta$ in to $M$ subcomponents, $\{\btheta_1, \ldots, \btheta_M\}$, we use the mean field variational Bayesian approximation to construct $q(\btheta)$, specifically $q(\btheta) = \prod_{m = 1}^M q_m(\btheta_m)$. This approach approximates the posterior as the product of these $q$-densities which are analogous to conditional posterior densities resulting from a Gibbs sampler \citep{Gelman2013}. Optimal densities are obtained iteratively with convergence (and therefore minimization) occurring when changes in the variation lower bound, $\up(\bY; q)$, become negligible. For a thorough introduction to variational Bayesian techniques with examples, see overviews by \cite{OrmerodWand2010} and \cite{Blei2017} or Chapter 13 of \cite{Gelman2013}. 
	
	The study of the frequentist properties of variational methods has seen growth in recent years as well. \cite{Bickel2013} and \cite{Zhang2020} discuss various asymptotic properties for mean-field stochastic blockmodels demonstrating their asymptotic normality rates and linear convergence rates in high dimensions, respectively. \cite{You2014} establish theoretical results for mean-field variational estimators in linear regression models showing they are consistent estimators. \cite{Wang2019} show the posterior converges to the KL minimizer of normal density that is centered at the truth. They also more generally establish the consistency of mean-field variational estimators, demonstrating that the variational expectation of the parameter is asymptotically normal. \cite{You2014} also derive a variational AIC (VAIC) and variational BIC (VBIC) for a general linear regression. The VAIC has good asymptotic properties and converges in probability to the standard AIC while the VBIC is equivalent to the standard BIC up to an $O(1)$ term. 
		
	There are, however, still limited statistical tools available for variational methods in various modeling classes, specifically with regard to model selection in variational mixed effects (VME) models. Starting with work by \cite{Harville1977} and \cite{LairdWare1982}, mixed effects models have a rich history of study. For more on these models in general, see the text by \cite{Fitz2011} or the review article by \cite{Laird2022}. In the variational context, they have also seen broad study by a number of authors including \cite{OrmerodWand2010,Ormerod2012,Tan2013,Yi2022,Menictas2023} and references therein. 
	While many of these methods allow an arbitrary number of random effects, doing so increases the parameter space of the model in each framework by $n + 1$ for every additional random effect that is added. That is, the number of parameters increases by the number of observations plus one for the corresponding variance component.
	
	For example, the algorithm described in \cite{OrmerodWand2010} requires a separate variance parameter estimation for each random effect in addition to the $n$ coefficients that get added to the model. The algorithm in these cases necessarily expands as more random effects are added. To accommodate a wide range of random effect structures in a single modeling class, our first step is to develop a parameter-efficient VME model for Gaussian outcomes which we layout in Section~\ref{s:vmes}. Using a B-spline basis expansion with one knot per subject, we reduce an arbitrarily large number of random effect coefficients to one per-subject. Thus, the size of the parameter space in need of estimation is the same as the random intercept only model. This streamlined model provides a framework for the inclusion of a wide range of random effect structures including multiple random intercepts, a random intercept and slope, and even random functional effects in a longitudinal scalar-on-function regression. From our parameter-efficient VME model, we next develop a criterion to select the random effect structure which, in turn, determines the covariance structure of the repeated samples---for details on this relationship, see \cite{Fitz2011}, Chapters 7 and 8.
	
	Akaike information criterion (AIC)-like metrics which can be used in REML-based estimation for covariance selection are limited, generally, to the linear regression case \citep{You2014}. A researcher working with longitudinal or repeat measures data would need to re-fit a variational model in a fully Bayesian framework to use the Watanabe-Akaike information criterion or deviance information criterion to perform model selection \citep[Chapter 7]{Gelman2013}. This is inefficient and defeats one of the purposes of using variational methods in the first place: the computational gains they afford over fully Bayesian implementations while still providing consistent estimators \citep{Wang2019}. To address this, we build off of the work by \cite{You2014} to derive (Section~\ref{s:vaic}) and study (Section~\ref{s:sim}) a VAIC for the parameter-efficient VME model.
	
	We explore the properties of our VAIC in a simulation that tests its ability to correctly identify ``true'' random effect structures and, consequently, identify the covariance structure (Section~\ref{s:sim}). Our VAIC performs well in correctly identifying the random effect structure for both more traditional random intercept and random slope models as well as for selection in longitudinal scalar-on-function regression where the random effect can potentially be a functional effect. To illustrate the use of our VAIC in practice, we apply the method to two data sets: a lead level study and a diffusion tensor image (DTI) study (Section~\ref{s:app}). Finally, we provide a discussion of the method in Section~\ref{s:disc}
	
\section{Parameter-Efficient VME Models}
\label{s:vmes}

Let $Y_{ij}$ denote the $j$th response of the $i$th subject. We define the following models:
	\begin{align}
		Y_{ij} &= \bx_{ij}'\bbeta + u_i + \varepsilon_{ij} \text{ (random intercept) and} \label{eq:ri}\\
		Y_{ij} &= \bx_{ij}'\bbeta + \bz_{ij}\bb_{i}  + \varepsilon_{ij}\text{ (general random effects)},\label{eq:gre}
	\end{align}
where $u_i$ is a random intercept, $\bb_{i}$ is a vector of random coefficients, and $\varepsilon_{ij}$ is the model error. For all models, we assume iid Gaussian within-subject errors, $\varepsilon_{ij} \stackrel{iid}{\sim} N(0, \sigma_e^2)$. The fixed effects, $\bx_{ij}$, are a $1\times P$ vector depending on the number of covariates and $\bz_{ij}$ is a $1\times Q$ vector with $Q$ depending on the dimensionality of the general random effects structure. For example, when using a random intercept plus a random slope, $Q = 2$ since $\bz_{ij} = [\begin{array}{cc} 1 & x_{p,ij}\end{array}]$ for some covariate $x_{p,ij}$. The vectors $\bbeta$ and $\bb_i$ are then $P\times 1$ and $Q \times 1$, respectively. Hierarchical data structures might admit random effects structures of the form $\bz_{ij} = [\begin{array}{ccc} x_{1,ij} & x_{2,ij} & x_{3,ij}\end{array}]$  where $x_{1,ij}, x_{2,ij},$ and $x_{3,ij}$ indicate membership in potentially overlapping groups. Data with this structure could be clustered on the municipality, county, and state level. In this case, the vector $\bb_i$ consists of three different random intercepts. More general forms of $\bz_{ij}$ require additional considerations.

If a functional predictor is of interest, we may wish to incorporate it into the random effect structure or simply use a random intercept. Such models can be written as
	\begin{align}
		Y_{ij} &= \bx_{ij}'\bbeta + \int_{t\in\mathcal{T}} w_{ij}(t)\gamma(t) dt + b_i + \varepsilon_{ij} \text{ (random intercept) and} \label{eq:rif}\\
		Y_{ij} &= \bx_{ij}'\bbeta + \int_{t\in\mathcal{T}} w_{ij}(t)\gamma(t) dt + \int_{t\in\mathcal{T}} w_{ij}(t)\gamma_i(t) dt + \varepsilon_{ij}\text{ (random function)},\label{eq:rf}
	\end{align}
	where $w_{ij}(t)$ is a functional predictor observed on the domain $\mathcal{T}$. The functions $\gamma(t)$ and $\gamma_i(t)$ are the corresponding population averaged effect and subject-specific, or random, effect of $w_{ij}(t)$ over time. Functional data typically comes sampled on a grid, $\{t : t = t_1, \ldots t_T \}$ for $T$ total measurements. The grid need not be equally spaced, although without-loss-of-generality we assume it is. Equations~\eqref{eq:rif} and~\eqref{eq:rf} can then be expressed in terms of a matrix approximation to the integrals:
	\begin{align*}
		Y_{ij} &= \bx_{ij}'\bbeta + \bw_{ij}\bgamma + b_i + \varepsilon_{ij} \text{ (random intercept) and} \\
		Y_{ij} &= \bx_{ij}'\bbeta + \bw_{ij}\bgamma + \bw_{ij}\bgamma_i + \varepsilon_{ij}\text{ (random function)}.
	\end{align*}
When viewed like this, we see that Equations~\eqref{eq:rif} and~\eqref{eq:rf} are really just special cases of Equations~\eqref{eq:ri} and~\eqref{eq:gre} with $\bw_{ij}\bgamma$ incorporated into the fixed effects and the general random effect structure specified to be $\bw_{ij}\bgamma_i$.

Stacking the scalar longitudinal measurements into a vector, the fixed effect vectors into matrices, and the random effect vectors into matrices, we can represent Equations~\eqref{eq:ri} to~\eqref{eq:rf} in a general matrix form:
\begin{align}
	\bY = \bX\bbeta + \bZ_u\bU + \bveps\text{ (random intercept) and} \label{eq:riMat}\\
	\bY = \bX\bbeta + \bZ_b\bB + \bveps \text{ (general random effects)}.\label{eq:grMat}
\end{align}
Let $N = \sum_{i=1}^n m_i$ for $m_i$ observations per subject with $n$ total subjects. Then, $\bY$ is $N\times1$, $\bX$ is $N \times P$, $\bZ_u$ is $N\times n$, and $\bZ_b$ is $N\times nQ$. The vectors $\bU$ and $\bB$ are $n\times1$ and $nQ\times 1$, respectively. As is common for subject-specific intercepts, we we will impose a ridge-type penalty prior via the mixed model formulation of penalized regression \citep{Ruppert2003}. We will penalize $\bB$ as well but when $Q$ is large, the dimensionality of $\bB$ can get unruly, particularly for large $n$. Thus, in addition to the usual ridge-type penalty prior, we propose first using a basis expansion with knots pre-selected to keep the number of additional coefficients that need to be estimated at $n$. This effectively reduces the general random effects model to a modeling context similar to the random intercept in terms of parameter space where just one subject-specific coefficient needs to be estimated.

To achieve this reduction, let $\bTheta$ be an $nQ \times n$ matrix of known basis functions. Then, $\bB = \bTheta\bB^{*}$ where $\bB^*$ is $n\times 1$. Substituting into Equation~\eqref{eq:grMat}, the general random effects model becomes
	$\bY = \bX\bbeta + \bZ_b\bTheta\bB^{*} + \bveps$,
where both $\bZ_b$ and $\bTheta$ are known. We take $\bTheta$ to be the popular B-spline basis functions but any basis expansion that can be limited to $n$ basis coefficients could work. As is common with B-splines, we penalize their fit using an appropriate penalty matrix, see \cite{Eilers1996} for additional details. The specific penalty matrix is $\mathcal{P} = \xi D_0 + (1-\xi)D_2$ which is weighted between the zeroth derivative matrix ($D_0$) and the second derivative matrix ($D_2$). The parameter $\xi \in [0, 1]$ controls the desired tradeoff between shrinkage and smoothness, with values near 0 favoring shrinkage. When incorporating a fixed functional effect, we also perform a basis expansion using B-splines and penalize their fit, similar to \cite{Goldsmith2012}. The matrix form of such a model might be $\bY = \bX\bbeta + \bW\bXi\bgamma^* + \bZ_b\bTheta\bB^{*} + \bveps$ where $\bW$ stacks the $\bw_{ij}$ vectors into a functional fixed effect matrix and $\bXi$ is a $T\times K_{\gamma}$ matrix of B-spline basis functions.

We perform this basis expansion for any general random effect structure beyond the random intercept only model. Depending on the size of $Q$ and structure of $\bZ_b$, $\xi$ may be close to 0 or 1. When a random intercept and slope model is used, we select $\xi = 0.99$ since the function is linear. But when a random function is used, we take $\xi = 0.01$ to induce shrinkage. The degree of the B-splines must also take into account the structure of $\bZ_b$. Thus for random intercept and slope models, we use a linear (degree 1) B-spline but for random function models, we use cubic B-splines. Via the mixed model representation of penalized regression in the Bayesian framework, the penalty is applied via a specific prior.

Penalized model components include $\bgamma$ under the longitudinal scalar-on-function regression model and $\bB^*$ for the general random effect structure. The penalty priors on $\bgamma$ and $\bB^*$ are $\bgamma \sim N(\textbf{0}, \lambda_{\gamma} \mathcal{P}_{\gamma}^{-1})$ and $\bB \sim N(\textbf{0}, \lambda_{B} \mathcal{P}_{B}^{-1})$, where $\lambda_{\gamma}$ and $\lambda_{B}$ are tuning-parameters for corresponding penalty matrices, $\mathcal{P}_{\gamma}$ and $\mathcal{P}_{B}$---both of form $\mathcal{P}$. For the functional fixed effect, we select a small number of knots, setting $K_{\gamma} = 8$ and set $\xi = 0.01$. Regardless of the model, we place weakly informative priors on the components of $\bbeta$, $\beta_p \stackrel{iid}{\sim} N(0, \sigma_b^2)$ with $\sigma_b^2$ fixed at something large ($\sigma_b^2 = 1000$). For the random intercept model, the vector $\bU$ consists of the $b_i$ which are independent normals, $b_i \stackrel{iid}{\sim} N(0, \sigma_u^2)$. We place inverse-gamma priors (and hyper-priors) on all variance components: $\sigma^2 \sim IG(a_e, b_e)$,  $\lambda_{\gamma} \sim IG(a_{\gamma}, b_{\gamma})$, $\lambda_{B} \sim IG(a_{B}, b_{B})$, and $\sigma_u^2 \sim IG(a_u, b_u)$. The hyper-parameters for each variance prior are set to something small, $0.01$ for example.

Using a mean field variational Bayesian approximation, we obtain approximation densities or $q$-densities for each component of the model. We let $\btheta$ generically denote the coefficients which vary by model. For non-functional fixed effects, the random intercept only model has mean parameters $\btheta = [\begin{array}{cc} \bbeta & \bU \end{array}]$ while the general random effect model has $\btheta = [\begin{array}{cc} \bbeta & \bB^* \end{array}]$. When using a functional predictor, the random intercept only model has mean parameters $\btheta = [\begin{array}{ccc} \bbeta & \bgamma^* & \bU \end{array}]$ and the random function model has $\btheta = [\begin{array}{ccc} \bbeta & \bgamma^* & \bB^* \end{array}]$. The $q$-densities for all possible model parameters are $q(\btheta) \sim N\left[\bmu_{q(\btheta)}, \bSigma_{q(\btheta)}\right]$, $q(\sigma^2) \sim IG\left[ a_e + \frac{N}{2}, B_{q(\sigma^2)} \right]$, $q(\lambda_B) \sim IG\left[ a_B + \frac{K_B}{2}, B_{q(\lambda_B)}\right]$, $q(\sigma^2_u) \sim IG\left[ a_U + \frac{K_U}{2}, B_{q(\sigma^2_u)}\right]$, and $q(\lambda_{\gamma}) \sim IG\left[ a_{\gamma} + \frac{K_{\gamma}}{2}, B_{q(\lambda_{\gamma})}\right]$. The subscript-${q(\cdot)}$ notation indicates the parameter to which the quantity belongs under the mean field approximation.

\begin{algorithm}
\caption{Variational algorithm for the model with a non-functional fixed effect. The algorithm for the model with the random intercept replaces $B_{q(\lambda_B)}$ with $B_{q(\sigma_u^2)}$, $\bB^*$ with $\bU$, $a_B$ with $a_U$, $b_B$ with $B_U$, and $K_B$ with $K_U$. The penalty matrix  $\mathcal{P}_B$, is set to $I_{n\times n}$. $\Delta \log[ \up(\bY; q) ]$ denotes the change in $\log[ \up(\by; q) ]$.}\label{a:ri}
\begin{algorithmic}
\Require $B_{q(\sigma^2)}, B_{q(\lambda_B)} > 0$, and $\epsilon > 0$, small
	\While{$\Delta \log[ \up(\bY; q) ] > \epsilon$}
		\State $\bSigma_{q(\btheta)} \gets \left[ \frac{a_e + \frac{N}{2}}{B_{q(\sigma^2)}}\bC'\bC + \text{blockdiag}\left\{ (\sigma_b^2)^{-1} I_{p\times p}, \frac{a_B + \frac{1}{2} K_B}{B_{q(\lambda_B)}} \mathcal{P}_B  \right\}\right]^{-1}$
		\State $\bmu_{q(\btheta)} \gets \left( \frac{a_e + \frac{N}{2}}{B_{q(\sigma^2)}}\right)\bSigma_{q(\btheta)}\bC'\bY $ 
		\State  $B_{q(\sigma^2)} \gets b_e + \frac{1}{2}\left[ \left\{\bY - \bC\bmu_{q(\btheta)} \right\}'\left\{\bY - \bC\bmu_{q(\btheta)} \right\} + \text{tr}\left\{ \bC'\bC \right\}\bSigma_{q(\btheta)} \right]$
		\State $B_{q(\lambda_B)} \gets b_ B + \frac{1}{2}\left[ {\bmu_{q(\bB^*)}}'\bmu_{q(\bB^*)} + \text{tr}\left\{\frac{a_B + \frac{1}{2} K_B}{B_{q(\sigma_B^2)}}  \mathcal{P}_B \right\} \right]$
	\EndWhile
\end{algorithmic}
\end{algorithm}

Algorithms~\ref{a:ri} and~\ref{a:gr} present the variational algorithms for the parameter-efficient VMEs for non-functional and functional fixed effects, respectively. These algorithms are similar to those discussed in \cite{OrmerodWand2010} and illustrate our approach to estimating the general random effect models---a slight alteration of each algorithm produces the random intercept only models. The design matrix, $\bC$, described in each combines all mean-model components. Thus, for non-functional models, $\bC = [\begin{array}{ccc} \bX & \bZ_u \end{array}]$ under the random intercept only model and $\bC = [\begin{array}{ccc} \bX & \bZ_b\bTheta \end{array}]$ for the general random effects model. In the scalar-on-function models, $\bC = [\begin{array}{ccc} \bX & \bW\bXi & \bZ_u \end{array}]$ when using a random intercept only and $\bC = [\begin{array}{ccc} \bX & \bW\bXi & \bZ_b\bTheta \end{array}]$.

\begin{algorithm}
\caption{Variational algorithm for the model with a functional fixed effect.  The algorithm for the model with the random intercept replaces $B_{q(\lambda_B)}$ with $B_{q(\sigma_u^2)}$, $\bB^*$ with $\bU$, $a_B$ with $a_U$, $b_B$ with $B_U$, and $K_B$ with $K_U$. The penalty matrix  $\mathcal{P}_B$, is set to $I_{n\times n}$. $\Delta \log[ \up(\by; q) ]$ denotes the change in $\log[ \up(\by; q) ]$.}\label{a:gr}
\begin{algorithmic}
\Require $B_{q(\sigma^2)}, B_{q(\lambda_{\gamma})}, B_{q(\lambda_B)} > 0$ and $\epsilon > 0$, small
	\While{$\Delta \log[ \up(\by; q) ] > \epsilon$}
		\State $\bSigma_{q(\btheta)} \gets \left[ \frac{a_e + \frac{N}{2}}{B_{q(\sigma^2)}}\bC'\bC + \text{blockdiag}\left\{ (\sigma_b^2)^{-1} I_{p\times p}, \frac{a_{\gamma} + \frac{1}{2} K_{\gamma}}{B_{q(\lambda_{\gamma})}} \mathcal{P}_{\gamma}, \frac{a_B + \frac{1}{2} K_B}{B_{q(\lambda_B)}} \mathcal{P}_B  \right\}\right]^{-1}$
		\State $\bmu_{q(\btheta)} \gets \left( \frac{a_e + \frac{N}{2}}{B_{q(\sigma^2)}}\right)\bSigma_{q(\btheta)}\bC'\bY $ 
		\State  $B_{q(\sigma^2)} \gets b_e + \frac{1}{2}\left[ \left\{\bY - \bC\bmu_{q(\btheta)} \right\}'\left\{\bY - \bC\bmu_{q(\btheta)} \right\} + \text{tr}\left\{ \bC'\bC \right\}\bSigma_{q(\btheta)} \right]$
			\State $B_{q(\lambda_{\gamma})} \gets b_ {\gamma} + \frac{1}{2}\left[ {\bmu_{q(\bgamma^*)}}'\bmu_{q(\bgamma^*)} + \text{tr}\left\{\frac{a_{\gamma} + \frac{1}{2} K_{\gamma}}{B_{q(\sigma_{\gamma}^2)}} \mathcal{P}_{\gamma} \right\} \right]$
		\State $B_{q(\lambda_B)} \gets b_ B + \frac{1}{2}\left[ {\bmu_{q(\bB^*)}}'\bmu_{q(\bB^*)} + \text{tr}\left\{\frac{a_B + \frac{1}{2} K_B}{B_{q(\sigma_B^2)}}  \mathcal{P}_B \right\} \right]$
	\EndWhile
\end{algorithmic}
\end{algorithm}

	Both algorithms iterate until changes in the variational lower bound become minimal. The $\log$ of this lower bound for the non-functional model in Algorithm~\ref{a:ri} is
	\begin{align*}
		\log[ \up(\bY; q) ] &= \frac{1}{2}(P + K_B) - \frac{N}{2}\log(2\pi) - \frac{P}{2}\log(\sigma_b^2) \\
			&+ \frac{1}{2}\log\left(|\bSigma_{q(\btheta)}|\right) - \frac{1}{2\sigma_b^2}\left[ {\bmu_{q(\bbeta)}}'\bmu_{q(\bbeta)} + \text{tr}\left\{\bSigma_{q(\bbeta)}\right\} \right] \\
			&- a_e\log(b_e) - \left(a_e + \frac{N}{2}\right)\log(B_{q(\sigma^2)}) + \log\left(\Gamma\left(a_e + \frac{N}{2}\right)\right) - \log\left(\Gamma(a_e)\right) \\
			&+ a_B\log(b_B) - \left(a_B + \frac{K_B}{2}\right)\log(B_{q(\lambda_B)}) + \log\left(\Gamma\left(a_B + \frac{K_B}{2}\right)\right) - \log(\Gamma(a_B)),
	\end{align*}
	where $\bmu_{q(\bbeta)}$ is $\bmu_{q(\btheta)}$ subset to the quantities corresponding to $\bbeta$, likewise for $\bSigma_{q(\bbeta)}$. For the random intercept model, the last line replaces with $a_B$ with $a_u$, $b_B$ with $b_u$, $K_B$ with $K_u$, and $\lambda_B$ with $\sigma^2_u$. The functional model has the lower bound above plus $a_{\gamma}\log(b_{\gamma}) - \left(a_{\gamma} + \frac{K_{\gamma}}{2}\right)\log(B_{q(\lambda_{\gamma})}) + \log\left(\Gamma\left(a_{\gamma} + \frac{K_{\gamma}}{2}\right)\right) - \log(\Gamma(a_{\gamma}))$. Code to implement all algorithms in \texttt{R} is available alongside this article and at \url{https://github.com/markjmeyer/VME}.

\section{Variational AIC}
\label{s:vaic}

\cite{You2014} propose a variational AIC (VAIC) for multiple linear regression which, as we discuss in Section~\ref{s:intro}, has good asymptotic properties and converges in probability to the standard AIC in that context. The general form of their VAIC is
		$VAIC \equiv - 2\log{p(\bY | \btheta^*)} + 2P^*_D$,
where $\btheta^* = E_q(\btheta)$ and $P^*_D = 2\log{p(\bY | \btheta^*)} - 2 E_q[\log{p(\bY|\btheta)}]$. The expectation is taken with respect to the variational approximation densities, i.e. the $q$-densities. We derive $\log{p(\bY | \btheta^*)}$ and $E_q[\log{p(\bY|\btheta)}]$ for the parameter-efficient VME to obtain
	\begin{align*}
		\log{p(\bY | \btheta^*)} = &-\frac{N}{2} \log(2\pi) -\frac{N}{2}\left[ \log\left\{B_{q(\sigma^2)}\right\} - \log\left(a_e +\frac{N}{2} - 1\right) \right] \\
		&- \frac{1}{2}\frac{a_e +\frac{N}{2} - 1}{ B_{q(\sigma^2)}}\left\{\bY - \bC\bmu_{q(\btheta)}\right\}'\left\{\bY - \bC\bmu_{q(\btheta)}\right\} \text{ and}\\
		E_q[ \log{p(y | \theta)} ] = &-\frac{N}{2} \log(2\pi) + \frac{N}{2} \left[ \psi\left(a_e +\frac{N}{2}\right) - \log\left\{ B_{q(\sigma^2)} \right\}   \right] \\
		&- \frac{1}{2}  \frac{a_e + \frac{N}{2}}{ B_{q(\sigma^2)} } \left[ \text{tr}\left\{ \bC\bSigma_{q(\btheta)} \bC' \right\} + \left\{\bY - \bC\bmu_{q(\btheta)} \right\}'\left\{ \bY - \bC\bmu_{q(\btheta)} \right\} \right],
	\end{align*}
	where $\psi(\cdot)$ denotes the digamma function and $\btheta$ depends on model specification. Combining, the VAIC for the parameter-efficient VME is then
	\begin{align*}
		VAIC &=  N\log\left(a_e +\frac{N}{2} - 1\right) + N \log(2\pi) - 2N \psi\left(a_e +\frac{N}{2}\right)  + N\log\left\{ B_{q(\sigma^2)} \right\}  \\
		&+ 2  \frac{a_e + \frac{N}{2}}{ B_{q(\sigma^2)} } \text{tr}\left\{ \bC\bSigma_{q(\btheta)} \bC' \right\} +   \frac{a_e + \frac{N}{2} + 1}{ B_{q(\sigma^2)} } \left\{\bY - \bC \bmu_{q(\btheta)} \right\}'\left\{\bY - \bC \bmu_{q(\btheta)} \right\}.
	\end{align*}
	A full derivation of this quantity is in the Supplementary Material. This formulation applies to both the random intercept and general random effect models in either the non-functional or scalar-on-function models. $\bSigma_{q(\btheta)}$ and $\bmu_{q(\btheta)}$ are the values from Algorithms~\ref{a:ri} and~\ref{a:gr} upon convergence and will depend on the model specification. Because the theoretical results in \cite{You2014} were derived for non-longitudinal models, we empirically examine the properties of the VAIC we derive for use in selecting the structure of the random effects in parameter-efficient VMEs.

As with the standard AIC, the smallest VAIC often suggests the best model fit. We refer to this as the minimum decision rule. Parsimonious models are, however, also desirable when model fits are similar. Thus other rules might impose a threshold on the absolute difference in VAICs between two candidate models. When the difference is under the threshold, the more parsimonious model (i.e., the random intercept model) would be selected. We examine the minimum decision rule along with absolute difference thresholds of varying tolerances in our simulation study. Code to implement VAICs for VMEs is also available online at  \url{https://github.com/markjmeyer/VME}.

\section{Simulation}
\label{s:sim}

Our empirical evaluation of the VAIC for VMEs considers both a more classic longitudinal covariance selection problem as well as a longitudinal scalar-on-function covariance selection. In the former, we aim to select between a model with a random intercept alone and one with both a random intercept and a random slope. In the latter, the selection is between a model with a random intercept alone and one with a random functional effect. For both selection problems, we assume a balanced design for sample sizes of $n = 20, 50, 100$ with number of repeated observations per subject of $m_i = 2, 3, 4, 5$. The simulated data models are based, in part, off of the data illustrations. Thus for the classic longitudinal case, the fixed effects model is $\beta_0 + \beta_1\text{treatment} + \beta_2\text{time}$ with the true values of $\beta_0, \beta_1,$ and $\beta_2$ set to 25, $-5$, and $-1$, respectively. These values were based off of our analysis of the lead level data. The covariate Treatment is binary while Time is continuous. In the scalar-on-function longitudinal simulation, the mean model is $\beta_0 + \bw_{ij}\bgamma$ where $\beta_0$ is set to 50 to mimic the DTI data. For $ \bw_{ij}$ and $\bgamma$, we take $T = 50, 100$ and consider three true values for $\bgamma$: cyclical, peak, and sigmoidal. Graphs of these curves, along with the equations to generate them, are in the Supplementary Material.

For each combination of settings, we generate 500 simulated datasets and apply both a correctly specified and misspecified model to the data. In general, every simulated dataset is analyzed with a random intercept alone and a general random effect which varies depending on whether the setting is functional or not. We determine the VAIC for every model as well as the bias and mean squared error (MSE) of the mean model---in the case of the scalar-on-function simulation, we find mean integrated squared error (MISE). 
When evaluating the discrimination of the VAIC, we examine four different decision rules: the minimum ($\min$) rule where the smallest model is selected, the under two ($< 2$) rule where the more parsimonious model is selected when the absolute value of the difference in two VAICs is within 2 and the under five ($< 5$) and under ten ($<10$) rules which are the same as the under two rule but with thresholds of five and ten. The more parsimonious model is always the random intercept only model.

\begin{table}
\centering
\caption{Percent of models correctly identified, by decision rule, for random intercept only vs. random intercept $+$ slope models averaging over $m_i$. Int. is short for intercept.\label{t:corIS}}
\begin{tabular}{lrcccc}
  \hline
\multirow{2}{*}{True Model} & \multirow{2}{*}{$n$}  & \multicolumn{4}{c}{Decision Rule}  \\
\cline{3-6}
  &  &  $\min$ & $< 2$ & $< 5$ & $< 10$ \\ 
  \hline
 Int. Only & 20 & 96.10\% & 97.15\% & 98.45\% & 99.25\% \\  
       & 50  & 99.55\% & 99.70\% & 99.85\% & 99.95\% \\  
       & 100  & 99.95\% & 99.95\% & 99.95\% & 100.0\% \\  
   \cline{2-6}
 Int. + Slope  & 20 & 99.00\% & 98.75\% & 98.45\% & 98.10\% \\   
      & 50 & 99.90\% & 99.85\% & 99.85\% & 99.85\%  \\  
       & 100  & 100.0\% & 100.0\% & 100.0\% & 100.0\% \\  
    \hline
\end{tabular}
\end{table}

\begin{table}
\centering
\caption{Percent of models correctly identified, by decision rule and averaging over $m_i$, for random intercept only vs. random functional effect when $T = 50$. Int. is short for intercept, Func. denotes function.\label{t:cort50}}
\begin{tabular}{llrcccc}
  \hline
\multirow{2}{*}{Effect} & \multirow{2}{*}{True Model} & \multirow{2}{*}{$n$}  & \multicolumn{4}{c}{Decision Rule}  \\
\cline{4-7}
  &  &  & $\min$ & $< 2$ & $< 5$ & $< 10$ \\ 
  \hline
Cyclical  & Int. & 20 & 99.30\% & 99.60\% & 99.70\% & 99.80\%  \\  
      &  & 50  & 100.0\% & 100.0\% & 100.0\% & 100.0\%  \\  
      &  & 100  & 100.0\% & 100.0\% & 100.0\% & 100.0\% \\  
   \cline{3-7}
 & Func.  & 20 & 100.0\% & 100.0\% & 100.0\% & 100.0\%  \\  
      &  & 50  & 100.0\% & 100.0\% & 100.0\% & 100.0\%  \\  
      &  & 100  & 100.0\% & 100.0\% & 100.0\% & 100.0\% \\  
   \cline{2-7}
Peak & Int. & 20 & 99.30\% & 99.45\% & 99.65\% & 99.80\%  \\  
      &  & 50  & 100.0\% & 100.0\% & 100.0\% & 100.0\% \\  
      &  & 100  & 100.0\% & 100.0\% & 100.0\% & 100.0\% \\  
   \cline{3-7}
 & Func.  & 20 & 100.0\% & 100.0\% & 100.0\% & 100.0\% \\  
      &  & 50  & 100.0\% & 100.0\% & 100.0\% & 100.0\% \\  
      &  & 100  & 100.0\% & 100.0\% & 100.0\% & 100.0\% \\  
   \cline{2-7}
Sigmoidal & Int. & 20 & 99.30\% & 99.60\% & 99.70\% & 99.80\% \\  
      &  & 50  & 100.0\% & 100.0\% & 100.0\% & 100.0\% \\  
      &  & 100  & 100.0\% & 100.0\% & 100.0\% & 100.0\% \\  
   \cline{3-7}
 & Func.  & 20 & 100.0\% & 100.0\% & 100.0\% & 100.0\% \\  
      &  & 50  & 100.0\% & 100.0\% & 100.0\% & 100.0\% \\  
      &  & 100  & 100.0\% & 100.0\% & 100.0\% & 100.0\% \\  
   \hline
\end{tabular}
\end{table}

Tables~\ref{t:corIS} and~\ref{t:cort50} contain the percent of correctly identified models, by decision rule, for the classic longitudinal regression simulation and the $T = 50$ scalar-on-function regression simulation, respectively. The four rightmost columns show the percent of simulations where the true model was correctly identified by decision rule. The decision rules increasingly favor the parsimonious, i.e. random intercept model, moving left to right. Table~\ref{t:corIS} is broken down by the true model and number of subjects while Table~\ref{t:cort50} is further broken down by true curve. Both tables average over $m_i$ as the number of repeat samples did not impact the VAIC's discrimination. Under either model type, we see the VAIC correctly identifies the true model in excess of 96\% of time with many settings having correct specification in 100\% of simulations. Discrimination improves, insofar as it can, when true model is the random intercept only as the decision rule's threshold increases. There is a slight decrease when the true model is a random intercept and slope as the threshold increases, but it is minor and diminishes as the sample size increases. For the scalar-on-function model, the VAIC is quite good at selecting the correct model, regardless of true model, sample size, or decision rule (Table~\ref{t:cort50}). Similar results for the $T = 100$ case are in the Supplementary Material.

\begin{table}
\centering
\caption{Average bias and MSE for random intercept only vs. random intercept $+$ slope models averaging over $m_i$. Int. is short for intercept.\label{t:bmIS}}
\begin{tabular}{lrccccc}
  \hline
   & & \multicolumn{5}{c}{Specified Model} \\
 \multirow{2}{*}{True Model} & \multirow{2}{*}{$n$}  & \multicolumn{2}{c}{Average Bias} & & \multicolumn{2}{c}{MSE} \\
\cline{3-4} \cline{6-7}
     &  & Int. Only & Int. + Slope  &  & Int. Only & Int. + Slope \\ 
  \hline
   Int. Only  & 20 & 0.0012 & 0.0001 & & 0.1420  &  0.1920 \\  
        & 50 & 0.0017  &  0.0012 &  & 0.0562  &  0.0792 \\  
        & 100 & 0.0008 & $-$0.0001 & & 0.0278  &  0.0368 \\  
   \cline{2-7}
  Int. + Slope  & 20 & $-$0.0092  &  $-$0.0017 & & 0.9725  &  0.2131  \\   
       & 50 & $-$0.0004  &  0.0022 & & 0.3982  &  0.0794 \\  
       & 100 & $-$0.0064  &  $-$0.0021 & & 0.2052  &  0.0401 \\  
   \hline
\end{tabular}
\end{table}

Bias and MSE for the classic longitudinal model are in Table~\ref{t:bmIS} while Table~\ref{t:bmT50} contains the bias and MISE for the scalar-on-function model when $T = 50$. The values in each of these tables are averaged over all simulated datasets and over all $m_i$ since changes in $m_i$ did not impact these values. Both tables report the bias and MSE/MISE for the correctly specified as well as misspecified models with the columns indicating the specified model and the rows indicating the true model. On average, bias tends to be similar for both correctly specified and misspecified models under the classic longitudinal regression setting. In the more complicated longitudinal scalar-on-function simulation, bias tends to be smaller on average when the model is correctly specified. Sample size does not appear to impact the bias with all settings having estimates of similar orders of magnitude. Bias results similar to those in Table~\ref{t:bmT50} for the case where $T = 100$ are in the Supplementary Material. Changing the sampling density does not alter the behavior of the model bias in simulation.

\begin{table}
\centering
\caption{Mean integrated bias and MISE for random intercept only vs. random functional effect when $T = 50$, averaging over $m_i$. Int. is short for intercept, Func. denotes function.\label{t:bmT50}}
\begin{tabular}{llrccccc}
  \hline
  & & & \multicolumn{5}{c}{Specified Model} \\
\multirow{2}{*}{Effect} & \multirow{2}{*}{True Model} & \multirow{2}{*}{$n$}  & \multicolumn{2}{c}{Average Bias} & & \multicolumn{2}{c}{MISE} \\
\cline{4-5} \cline{7-8}
   &  &  & Int. Only & Func.  &  & Int. Only & Func. \\ 
  \hline
Cyclical & Int. Only  & 20 & 0.0002 & $-$0.0004 & & 0.0032  &  0.0067 \\  
      &  & 50 & $-$0.0003 & 0.0001 & & 0.0011  &  0.0023 \\  
      &  & 100 & 0.0001 & 0.0005 & & 0.0006 &  0.0012 \\  
   \cline{3-8}
 & Func.  & 20 & 0.0063  &  0.0051 & & 0.1296  &  0.0496 \\  
      &  & 50 & 0.0101  &  0.0061 & & 0.0605  &  0.0205 \\  
      &  & 100 & 0.0032  &  0.0024 & & 0.0311  &  0.0099 \\  
      \cline{2-8}
Peak  & Int. Only  & 20 & $-$0.0026  &  $-$0.0207 & & 0.0043  &  0.0082 \\  
      &  & 50 & 0.0008 & $-$0.0076 & & 0.0021  &  0.0034 \\  
      &  & 100 & 0.0018  &  $-$0.0034 & & 0.0016  &  0.0022 \\  
   \cline{3-8}
 & Func.  & 20 & 0.0020  &  $-$0.0253 & & 0.1301  &  0.0501 \\  
      &  & 50 &  0.0086  &  $-$0.0071 & & 0.0613  &  0.0213 \\  
      &  & 100 & 0.0033  &  $-$0.0046 & & 0.0321  &  0.0109 \\  
   \cline{2-8}
Sigmoidal & Int. Only  & 20 & 0.0001 & $-$0.0252 & & 0.0032  &  0.0076 \\  
      &  & 50 & $-$0.0003 & $-$0.0093 & & 0.0011  &  0.0024 \\  
      &  & 100 & 0.0001 & $-$0.0042 & & 0.0006 & 0.0012 \\  
   \cline{3-8}
 & Func.  & 20 & 0.0037  &  $-$0.0397 & & 0.1295  &  0.0544\\  
      &  & 50 &  0.0085  &  $-$0.0133 & & 0.0605  &  0.0214 \\  
      &  & 100 & 0.0024  &  $-$0.0077 & & 0.0311  &  0.0102 \\  
   \hline
\end{tabular}
\end{table}

Bias does not account for the model variance but since our model selection focuses on identifying the covariance structure, MSE and MISE should be more informative metrics. This bears out as we see a much cleaner picture from these metrics: the MSEs and MISEs are lower for correctly specified models across all simulated datasets. It is particularly pronounced when misspecifying the models as intercept only. Correctly specified random intercept $+$ slope and random function models have MSEs/MISEs that are often an order of magnitude lower than the incorrectly specified model. Increasing sample sample size does tend to result in decreasing MSE and MISE. For results from the simulation with a sampling density of $T = 100$, see the Supplementary Material. Changing the sampling density does partially impact the MISE with some values nearly halved from what we present in Table~\ref{t:bmT50}.

\section{Data Illustrations}
\label{s:app}

Here we present illustrations of the use of the VAIC for model selection on two different data settings. The first, the Lead Level Study, is a more classic longitudinal analysis where the random effects structure could be a random intercept or random intercept $+$ slope. The second, the DTI Study, is a longitudinal study with a functional predictor and therefore a potentially functional random effects structure. The VAICs for all models are in Table~\ref{t:apps}. Additional results including estimated fixed effects from both studies and estimated curves from the DTI study are in the Supplementary Material. From these supplmental results, the efficiency of the method is also evident with the models converging within 25 iterations.

\begin{table}
\centering
\caption{VAICs for Data Illustrations. CCA denotes the fractional anisotropy profiles from the corpus callosum while RCST is the parallel diffusivity profiles from the right corticospinal tract. These are functional predictors in the mean model. The smallest VAIC per row is bolded. \label{t:apps}}
\begin{tabular}{llcc}
\hline
	 & & \multicolumn{2}{c}{Model} \\
	 \cline{3-4}
	Study & Mean Model & Random Intercept & General Random Effect \\
\hline
	Lead Level & treatment $+$ weeks & 3065.60 & \bf 3033.95 \\
	\cline{2-4}
	DTI  & visit $+$ CCA & \bf 2439.42 & 2499.96 \\
	 & visit $+$ RCST & \bf 2440.23 & 2494.48 \\
\hline
\end{tabular}
\end{table}

\subsection{Lead Level Study}

The Treatment of Lead-Exposed Children (TLC) Trial Group examined the effects of chelation treatment with the drug succimer on lead levels in children over time \citep{TLC2000}. The participants were toddlers aged 12 to 33 months who presented with blood lead levels between 20 and 44 $\mu$g/dL. A subset of this data is discussed in \cite{Fitz2011} and we use this subset as our first illustration. Measurements were taken at baseline, one week, four weeks, and six weeks. There are 100 children available for analysis, half of whom were randomly assigned to succimer and half to treatment with placebo. Each child has four measurements. The goal of this analysis is to select a random effects structure between a random intercept only model and a random intercept $+$ slope model.

From Table~\ref{t:apps}, we see the random intercept model returns a VAIC of 3065.60 while the more general model with a random intercept $+$ slope gives a VAIC of 3033.95. The fixed component of the mean model is the same for both models and set to the binary variable indicating treatment plus the variable weeks (coded 0, 1, 4, 6) which is treated as a continuous covariate. The absolute difference between the two VAIC is large, 31.65. Thus, under any of the decision rules described in Section~\ref{s:sim}, we'd select the model with the smallest VAIC which is the random intercept $+$ slope model. With the random component in place, one could select the fixed effect using VAIC as well, keeping the selected random effect structure constant. For example, an analysis of response profiles comparing children treated with succimer to those who received placebo might be appropriate to see if treatment effects vary over time \citep{Fitz2011}. 

\subsection{DTI Study}

\cite{Goldsmith2012} and \cite{Swihart2014} describe a longitudinal study examining the relationship between a functional predictors, diffusion tensor imaging (DTI) profiles, and the Paced Auditory Serial Addition Test (PASAT) score in patients with multiple sclerosis (MS). DTI measures the diffusivity of water in the white matter in the brain and can be used to generate images of the white matter. From these images, one can obtain summaries on a continuous scale of white matter tracts which are known as tract profiles. These are measured as functions of the distance along the tract. These tracts can be used to monitor disease progression over time. The PASAT score measures cognitive function, thus the goal of the DTI study in general is to see if the imaging tracts are predictive of changes in cognitive function. The goal of our analysis in this illustration is to use VAIC to select between a random intercept model and random function model for two types of tract profiles.

The two functional predictors of interest are the fractional anisotropy profiles from the corpus callosum (CCA) and the parallel diffusivity profiles from the right corticospinal tract (RCST). We examine models for these predictors separately. Consistent with the analysis in \cite{Goldsmith2012}, one scalar predictor is included which is a binary variable indicating two visits (coded 0) or more than two (coded 1). In total, there are 100 MS patients available for the analysis with between two and eight measurements (2.5 visits per subject on average). The CCA profiles contain 93 measurements while the RCST profiles have 53. When the fixed component of the model is visit $+$ CCA, the random intercept model gives a VAIC of 2439.42 while the random function model returns a VAIC of 2499.96. When the functional effect is RCST and the fixed component is visit $+$ RCST, the random intercept model has a VAIC of 2440.23 while the random functional model has a VAIC of 2494.48. The absolute differences in the VAIC are 60.54 and 54.25, respectively. Thus for both functional predictors, the simpler random effect of the random intercept only model structure is preferable. The DTI data is publicly available in the \texttt{refund} \texttt{R} package \citep{refund2022}.


\section{Discussion}
\label{s:disc}

In this manuscript, we derive a VAIC for VME models under random intercept and general random effects structures. Our simulation demonstrates the discriminatory properties of the VAIC which selects the correct random effect structure in excess of 96\% of the time across all scenarios. We show the benefit of selecting the correct random effect structure and its impact on bias and MSE/MISE where correctly specified models have lower MSE (or MISE) in particular. These results held under a range of sample sizes and under differing numbers of repeat measures. In the scalar-on-function case, the results also held up under different curves type. The VAIC is applicable to a range of VME models with varying levels of complexity in their random effect structures. Our data illustration further supports this with applications to a more classical longitudinal model in the Lead Level Study and to a longitudinal scalar-on-function regression setting in the DTI study. In both cases, the VAIC clearly suggests a preferable random effect structure which, once selected, can be used to further select the fixed effect structure and analyze the specific scientific questions of interest.

The VAIC performs well in the VME setting because in the Bayesian modeling context, the different random effect models constitute different mean models in the Frequentist sense. In other words, there is no difference in how the ``fixed'' and ``random'' effects are treated in the original Bayesian model. All unknown parameters are supplied with a prior so the distinction between ``fixed'' and ``random'' in a Bayesian sense reduces to a distinction between population-level and subject-level effects. We are ultimately selecting between different subject-specific mean models in the Frequentist sense and therefore benefit from the asymptotic properties of the VAIC derived by \cite{You2014}. This is evident from the results of our empirical study.

VMEs can include functional effects in a longitudinal scalar-on-function regression models which have seen extensive study in various contexts, see for example work by \cite{Goldsmith2012, Gertheiss2013, Swihart2014, Kundu2016, Staicu2020, Cui2022}, and references therein. To our knowledge, the previous approaches for longitudinal scalar-on-function do not consider VAIC-like model selection criteria. Our parameter-efficient approach fits within the penalized functional regression framework described by \cite{Goldsmith2012}, albeit with a more general and potentially functional random effect. But the VAIC we derive is not necessarily specific to this framework and is applicable to any variational Bayesian version of these longitudinal scalar-on-function methods. The generality of the VAIC calculation is due its reliance on the log of the likelihood. Thus, all that is required is a Gaussian model with variational estimates of $B_{q(\sigma^2)}$, $\bmu_{q(\btheta)}$, and $\bSigma_{q(\btheta)}$, as well as the data $\bY$ and $\bC$.

\bigskip

\begin{center}
{\large\bf ACKNOWLEDGEMENTS}
\end{center}

The MRI/DTI data were collected at Johns Hopkins University and the Kennedy-Krieger Institute.

\begin{center}
{\large\bf SUPPLEMENTARY MATERIAL}
\end{center}

\begin{description}

\item[Supplementary Material to Model Selection in Variational Mixed Effects Models:] Additional details and results referenced in Sections~\ref{s:vaic},~\ref{s:sim}, and~\ref{s:app}. (.pdf file)

\item[R-code for VAICs and VMEs:] R-code to implement the VAIC for VMEs as well as replicate the simulation and data illustration results are freely available online at \url{https://github.com/markjmeyer/VME}.

\item[Lead level data set:] The lead level data set used in the illustration of VAICs for VMEs in Section~\ref{s:app}. (.txt file)

\end{description}

%

\bibliographystyle{agsm}

\bibliography{fullbib}
\end{document}